\documentclass[preprint,showpacs,preprintnumbers]{revtex4}
\usepackage{amssymb}
\usepackage{amsmath}
\usepackage{graphicx}
\usepackage{dcolumn}
\usepackage{bm}

\input{tcilatex}
\begin{document}

\title{Electronic Band gaps and transport properties inside graphene
superlattices with one-dimensional periodic squared potentials}
\author{Li-Gang Wang$^{1,2}$ and Shi-Yao Zhu$^{1,2,3}$}
\date{2010.02.11}
\affiliation{$^{1}$Centre of Optical Sciences and Department of Physics, The Chinese
University of Hong Kong, Shatin, N. T., Hong Kong, China\\
$^{2}$Department of Physics, Zhejiang University, Hangzhou, 310027, China\\
$^{3}$Department of Physics, Hong Kong Baptist University, Kowloon Tong,
Hong Kong, China}

\begin{abstract}
The electronic transport properties and band structures for the
graphene-based one-dimensional (1D) superlattices with periodic squared
potentials are investigated. It is found that a new Dirac point is formed,
which is exactly located at the energy which corresponds to the zero
(volume) averaged wavenumber inside the 1D periodic potentials. The location
of such a new Dirac point is robust against variations in the lattice
constants, and it is only dependent on the ratio of potential widths. The
zero-averaged wavenumber gap associated with the new Dirac point is
insensitive to both the lattice constant and the structural disorder, and
the defect mode in the zero-averaged wavenumber gap is weakly dependent on
the insident angles of carriers.
\end{abstract}

\maketitle

\section{Introduction}

Recently, the experimental realization of a stable single layer of carbon
atoms densely packed in a honeycomb lattice has aroused considerable
interest in study of their electronic properties \cite%
{Novoselov2004,Zhang2005}. Such kind of material is well known as graphene,
and the low-energy charge carriers in pristine graphene are formally
described by a massless Dirac equation with many unusual properties near the
Dirac point where the valence band and conduction band touch each other \cite%
{Novoselov2004,Zhang2005,Novoselov2005a,Katsnelson2006,reviews}, such as the
linear energy dispersion, the chiral behaviour, ballistic conduction and
unusual quantum Hall effect \cite{Novoselov2005a,Purewal2006},
frequency-dependent conductivity \cite{Kuzmenko2008}, gate-tunable optical
transitions \cite{Wang2008}, and so on.

Most recently, there have been a number of interesting theoretical
invesitigations on the graphene supperlatices with periodic potential or
barrier structures, which can be generated by diffierent methods such as
electrostatic potentials \cite{Bai2007,Park2008b,Barbier2008,Park2008c} and
magnetic barriers \cite%
{RamezaniMasir2008,RamezaniMasir2009,DellAnna2009,Ghosh2009}. Sometimes
periodic arrays of corrugations \cite{Guinea2008,Isacsson2008} have also
been proposed as graphene superlattices. It is well known that the
supperlattices are very sucessful in controlling the electronic structures
of many conventional semiconducting materials (e.g. see Ref. \cite{Tsu2005}%
). In graphene-based superlattices, researchers have found that a
one-dimensional (1D) periodic-potential superlattice may result in the
strong anisotropy for the low-energy charge carriers' group velocities that
are reduced to zero in one direction but are unchanged in another \cite%
{Park2008b}. Furthermore, Brey and Fertig \cite{Brey2009} have shown that
such behavior of the anisotropy is a precursor to the formation of further
Dirac points in the electronic band structures and new zero energy states
are controlled by the parameters of the periodic potentials. Meanwhile, Park 
\textit{et al.} \cite{Park2008a} pointed out that new massless Dirac
fermions, which are absent in pristine graphene, could be generated when a
slowly varying periodic potential is applied to graphene; and they further
found the unusual properties of Landau levels and the quantum Hall effect
near these new Dirac fermions, which are adjustable by the superlattic
potential parameters \cite{Park2009}. Finally it should also be mentioned
that the electronic transmission and conductance through a graphene-based
Kronig-Penney potential have been recently studied \cite{Barbier2009} and
the tunable band gap could be obtained in graphene with a noncentrosymmetric
superlattice potential \cite{Tiwari2009}.

Graphene superlattices are not only of theoretical interest, but also have
been experimental realized. For example, superlattice patterns with
periodicity as small as 5 nm have been imprinted on graphene using the
electron-beam induced deposition \cite{Meyer2008}. Epitaxially grown
graphenes on metal (ruthenium or iridium) surfaces \cite%
{Marchini2007,Vazquezdeparga2008,Pan2008,Sutter2008,Martoccia2008,Coraux2008,Pletikosic2009}
also show superlattice patterns with several nanometers (about 3 nm to 10
nm) lattice period. Fabrication of periodically patterned gate electrodes is
another possible way of making the graphene-based superlattices.

Motived by these studies, in this paper, we will consider the robust
properties of the electronic bandgap structures and transport properties for
the graphene under the external periodic potentials by applying appropriate
gate voltages. Following the previous work \cite{Barbier2008}, we evaluate
the effects of the lattice constants, the angles of the incident charge
carriers, the structural disorders and the defect potentials on the
properties of electronic band structures and transmissions. It is found that
a new Dirac point is exactly located at the energy with the zero (volume)
averaged wavenumber inside the 1D periodic potentials, and the location of
such a new Dirac point is not dependent on lattice constants but dependent
on the ratio of potential widths; and the position of the associated
zero-averaged wavenumber gap near the new Dirac point is not only
independent of lattice constants but is also weakly dependent on the
incident angles. With the increasing of the lattice contants, the
zero-averaged wavenumber gap will open and close oscillationaly but the
center position of this gap does not depend on the lattice constants.
Furthermore it is shown that the zero-averaged wavenumber gap is insensitive
to the structural disorder, while the other opened gaps in 1D periodic
potentials are highly sensitive to the structural disorder. Finally we also
find that the defect mode inside the zero-averaged wavenumber gap is weakly
dependent on the insident angles while the defect mode in other gaps are
highly dependent on the incident angles.

The outline of this paper is the following. In Sec. II, with the help of the
additional two-component basis, we introduce a new transfer matrix method,
which is different from that in Ref. \cite{Barbier2008}, to calculate the
reflection, transmission, and the evolution of the wave function; our
transfer matrix method is very useful to deal with the periodic- or
multi-squared potentials. In Sec. III, the various effects of the lattice
constants, the incident angles of carriers, and the structural disorders on
the electronic band structures are discussed in detail; furthermore the
transport properties of the defect mode inside the zero-averaged wavenumber
gap are also discussed. Finally, in Sec. IV, we summarize our results and
draw our conclusions.

\section{Transfer Matrix method for the structures of periodic potentials in
the mono-layer graphene}

The Hamiltonian of an electron moving inside a mono-layer graphene in the
presence of the electrostatic potential $V(x)$, which only depends on the
coordinate $x$, is given by%
\begin{equation}
\hat{H}=v_{F}\mathbf{\sigma }\cdot \mathbf{p}+V(x)\hat{I},
\label{Hamiltonian}
\end{equation}%
where $\mathbf{p}=(p_{x},p_{y})=(-i\hbar \frac{\partial }{\partial x}%
,-i\hbar \frac{\partial }{\partial y})$ is the momentum operator with two
components, $\mathbf{\sigma }=(\sigma _{x},\sigma _{y})$, and $\sigma
_{x},\sigma _{y}$ are pauli matrices of the pseudospin, $\hat{I}$ is a $%
2\times 2$ unit matrix, and $v_{F}\approx 10^{6}$m/s is the Fermi velocity.
This Hamiltonian acts on a state expressed by a two-component pseudospinor $%
\Psi =(\tilde{\psi}_{A},\tilde{\psi}_{B})^{T},$ where $\tilde{\psi}_{A}$ and 
$\tilde{\psi}_{B}$ are the smooth enveloping functions for two triangular
sublattices in graphene. Due to the translation invariance in the $y$
direction, the wave functions $\tilde{\psi}_{A,B}(x,y)$ can be written as $%
\tilde{\psi}_{A,B}(x,y)=\psi _{A,B}(x)e^{ik_{y}y}.$ Therefore, from Eq. (\ref%
{Hamiltonian}), we obtain%
\begin{eqnarray}
\frac{d\psi _{A}}{dx}-k_{y}\psi _{A} &=&ik\psi _{B},  \label{SD1} \\
\frac{d\psi _{B}}{dx}+k_{y}\psi _{B} &=&ik\psi _{A},  \label{SD2}
\end{eqnarray}%
where $k=[E-V(x)]/\hbar v_{F}$ is the wavevector inside the potential $V(x)$%
, $E$ is the incident energy of a charge carrier, and $k_{0}=E/\hbar v_{F}$
corresponds to the incident wavevector. Obviously, when $E<V(x)$, the
wavevector inside the barrier is opposite to the direction of the electron's
velocity. This property leads to a Veselago lens in graphene $p-n$
junctions, which has been predicted by Cheianov, Fal'Ko and Altshuler \cite%
{Cheianov2007}.

In what follows, we assume that the potential $V(x)$ is comprised of
periodic structures of squared potentials as shown in Fig. 1. Inside the $j$
th potential, $V_{j}(x)$ is a constant, therefore, from Eqs. (\ref{SD1}) and
(\ref{SD2}), we can obtain%
\begin{eqnarray}
\frac{d^{2}\psi _{A}}{dx^{2}}+(k_{j}^{2}-k_{y}^{2})\psi _{A} &=&0,
\label{HoM1} \\
\frac{d^{2}\psi _{B}}{dx^{2}}+(k_{j}^{2}-k_{y}^{2})\psi _{B} &=&0.
\label{HoM2}
\end{eqnarray}%
Here the subscript "$j$" denotes the quantities in the $j$ th potential. The
solutions of Eqs. (\ref{HoM1}) and (\ref{HoM2}) are the following forms%
\begin{eqnarray}
\psi _{A}(x) &=&ae^{iq_{j}x}+be^{-iq_{j}x},  \label{SLA} \\
\psi _{B}(x) &=&ce^{iq_{j}x}+de^{-iq_{j}x},  \label{SLB}
\end{eqnarray}%
where $q_{j}=$sign$(k_{j})\sqrt{k_{j}^{2}-k_{y}^{2}}$ is the $x$ component
of the wavevector inside the $j$ th potential $V_{j}$ for $%
k_{j}^{2}>k_{y}^{2}$, otherwise $q_{j}=i\sqrt{k_{y}^{2}-k_{j}^{2}}$; and $a$
($c$) and $b$ ($d$) are the amplitudes of the forward and backward
propagating spinor components. Substituting Eqs. (\ref{SLA}) and (\ref{SLB})
into Eqs. (\ref{SD1}) and (\ref{SD2}), we can find the relations%
\begin{eqnarray}
c &=&\frac{ik_{j}}{iq_{j}+k_{y}}a,  \label{RL1} \\
d &=&-\frac{ik_{j}}{iq_{j}-k_{y}}b.  \label{RL2}
\end{eqnarray}%
Using Eqs. (\ref{RL1}) and (\ref{RL2}), we may obtain%
\begin{eqnarray}
\psi _{A}(x) &=&ae^{iq_{j}x}+be^{-iq_{j}x},  \label{SLN1} \\
\psi _{B}(x) &=&a\frac{ik_{j}}{iq_{j}+k_{y}}e^{iq_{j}x}-b\frac{ik_{j}}{%
iq_{j}-k_{y}}e^{-iq_{j}x}.  \label{SLN2}
\end{eqnarray}

In order to derive the connection for the wave functions $\psi _{A,B}(x)$
between any two positions $x_{j-1}$ and $x_{j-1}+\Delta x$ in the $j$ th
potential, we assume a basis $\mathbf{\Phi }(x)=\binom{\phi _{1}(x)}{\phi
_{2}(x)},$ which are expressed as,%
\begin{eqnarray}
\phi _{1}(x) &=&ae^{iq_{j}x}+be^{-iq_{j}x},  \label{bas1} \\
\phi _{2}(x) &=&ae^{iq_{j}x}-be^{-iq_{j}x}.  \label{bas2}
\end{eqnarray}%
Using the above basis, we can re-write Eqs. (\ref{SLN1}) and (\ref{SLN2}) as
the following form:%
\begin{equation}
\binom{\psi _{A}(x)}{\psi _{B}(x)}=R_{j}(E,k_{y})\binom{\phi _{1}(x)}{\phi
_{2}(x)},  \label{Pisi-phai}
\end{equation}%
where 
\begin{equation}
R_{j}(E,k_{y})=\left( 
\begin{array}{cc}
1 & 0 \\ 
i\sin \theta _{j} & \cos \theta _{j}%
\end{array}%
\right) .
\end{equation}%
Here $\theta _{j}=$arcsin($k_{y}/k_{j}$) is the angle between two components 
$q_{j}$ and $k_{y}$ in the $j$ th potential. Inside the same potential, from
the positions $x_{j-1}$ to $x_{j-1}+\Delta x$, the wavefunction $\binom{\psi
_{A}(x_{j-1})}{\psi _{B}(x_{j-1})}$ evolutes into another form $\binom{\psi
_{A}(x_{j-1}+\Delta x)}{\psi _{B}(x_{j-1}+\Delta x)},$ which can be also
expressed in terms of the above basis $\mathbf{\Phi }(x)$,%
\begin{equation}
\binom{\psi _{A}(x_{j-1}+\Delta x)}{\psi _{B}(x_{j-1}+\Delta x)}%
=T_{j}(\Delta x,E,k_{y})\binom{\phi _{1}(x_{j-1})}{\phi _{2}(x_{j-1})},
\end{equation}%
where%
\begin{equation}
T_{j}(\Delta x,E,k_{y})=\left( 
\begin{array}{cc}
\cos (q_{j}\Delta x) & i\sin (q_{j}\Delta x) \\ 
i\sin (q_{j}\Delta x+\theta _{j}) & \cos (q_{j}\Delta x+\theta _{j})%
\end{array}%
\right) .
\end{equation}%
Therefore, the relation between $\binom{\psi _{A}(x_{j-1})}{\psi
_{B}(x_{j-1})}$ and $\binom{\psi _{A}(x_{j-1}+\Delta x)}{\psi
_{B}(x_{j-1}+\Delta x)}\ $can be finially written as:%
\begin{equation}
\binom{\psi _{A}(x_{j-1}+\Delta x)}{\psi _{B}(x_{j-1}+\Delta x)}%
=M_{j}(\Delta x,E,k_{y})\binom{\psi _{A}(x_{j-1})}{\psi _{B}(x_{j-1})},
\end{equation}%
where the matrix $M_{j}$ is given by%
\begin{eqnarray}
M_{j}(\Delta x,E,k_{y}) &=&T_{j}(\Delta x,E,k_{y})R_{j}^{-1}(E,k_{y})  \notag
\\
&=&\left( 
\begin{array}{cc}
\frac{\cos (q_{j}\Delta x-\theta _{j})}{\cos \theta _{j}} & i\frac{\sin
(q_{j}\Delta x)}{\cos \theta _{j}} \\ 
i\frac{\sin (q_{j}\Delta x)}{\cos \theta _{j}} & \frac{\cos (q_{j}\Delta
x+\theta _{j})}{\cos \theta _{j}}%
\end{array}%
\right) .  \label{MMj2}
\end{eqnarray}%
It is easily to verify the equality: $\det [M_{j}]=1$. Here we would like to
point out that in the case of $E=V_{j}$, the transfer materix (\ref{MMj2})
should be recalculated with the similar method and it is given by 
\begin{equation}
M_{j}(\Delta x,E,k_{y})=\left( 
\begin{array}{cc}
\exp (k_{y}\Delta x) & 0 \\ 
0 & \exp (-k_{y}\Delta x)%
\end{array}%
\right) .  \label{MMj3}
\end{equation}

Meanwhile, in the $j$ th potential ($x_{j-1}<x<x_{j}$), the wave functions $%
\psi _{A,B}(x)$ can be also related with $\psi _{A,B}(x_{0})$ by 
\begin{equation}
\binom{\psi _{A}(x)}{\psi _{B}(x)}=Q(\Delta x_{j},E,k_{y})\binom{\psi
_{A}(x_{0})}{\psi _{B}(x_{0})},  \label{PropagationAB}
\end{equation}%
where $\Delta x_{j}=x-x_{j-1},$ $\psi _{A,B}(x_{0})$ are wave functions at
the incident end of the whole structure, and the matrix $Q$ is given by%
\begin{equation}
Q(\Delta x_{j},E,k_{y})=M_{j}(\Delta
x_{j},E,k_{y})\dprod\limits_{i=1}^{j-1}M_{i}(w_{i},E,k_{y}).  \label{QQQ}
\end{equation}%
Here $w_{i}$ is the width of the $i$ th potential, and the matrix $Q$ is
related to the transformation of the charge particle's transport in the $x$
direction. Thus we can know the wave functions $\psi _{A,B}(x)$ at any
position $x$ inside each potential with the help of a transfer matrix. The
initial two-component wave function $\binom{\psi _{A}(x_{0})}{\psi
_{B}(x_{0})}$ can be determined by matching the boundary condition. As shown
in Fig. 1, we assume that a free electron of energy $E$ is incident from the
region $x<0$ at any incident angle. In this region, the electronic wave
function is a superposition of the incident and reflective wavepackets, so
at the incident end ($x=0$), we have the functions $\psi _{A}(0)$ and $\psi
_{B}(0)$ as follows:%
\begin{eqnarray}
\psi _{A}(0) &=&\psi _{i}(E,k_{y})+\psi _{r}(E,k_{y})  \notag \\
&=&(1+r)\psi _{i}(E,k_{y}),  \label{PisiA0}
\end{eqnarray}%
where $\psi _{i}(E,k_{y})$ is the incident wavepacket of the electron at $x=0
$. In order to obtain the function $\psi _{B}(0)$ at the incident end, from
Eqs. (\ref{bas1}) and (\ref{bas2}), we can re-write the two-component basis
in terms of the incident wavepacket, which are given by 
\begin{eqnarray}
\phi _{1}(0) &=&\psi _{i}(E,k_{y})+\psi _{r}(E,k_{y})=(1+r)\psi
_{i}(E,k_{y}), \\
\phi _{2}(0) &=&\psi _{i}(E,k_{y})-\psi _{r}(E,k_{y})=(1-r)\psi
_{i}(E,k_{x}).
\end{eqnarray}%
Therefore, using the relation of Eq. (\ref{Pisi-phai}), we obtain%
\begin{eqnarray}
\psi _{B}(0) &=&i\sin \theta _{0}\phi _{1}(0)+\cos \theta _{0}\phi _{2}(0), 
\notag \\
&=&(e^{i\theta _{0}}-re^{-i\theta _{0}})\psi _{i}(E,k_{y}),  \label{PisiB0}
\end{eqnarray}%
where $\theta _{0}$ is the incident angle of the electron inside the
incident region ($x<0$). In the above derivations, we have used the relation 
$\psi _{r}(E,k_{y})=r\psi _{i}(E,k_{y})$, where $r$ is the reflection
coefficient. Obiviously, we have%
\begin{equation}
\left( 
\begin{array}{c}
\psi _{A}(0) \\ 
\psi _{B}(0)%
\end{array}%
\right) =\left( 
\begin{array}{c}
1+r \\ 
(e^{i\theta _{0}}-re^{-i\theta _{0}})%
\end{array}%
\right) \psi _{i}(E,k_{y}).  \label{WAVE00}
\end{equation}%
In the similar way, at the exit end we have%
\begin{equation}
\left( 
\begin{array}{c}
\psi _{A}(x_{e}) \\ 
\psi _{B}(x_{e})%
\end{array}%
\right) =\left( 
\begin{array}{c}
t \\ 
te^{i\theta _{e}}%
\end{array}%
\right) \psi _{i}(E,k_{y}),  \label{WAVEdd}
\end{equation}%
with the assumption of $\psi _{A}(x_{e})=t\psi _{i}(E,k_{y})$, where $t$ is
the transmission coefficient of the electronic wave function through the
whole structure, and $\theta _{e}$ is the exit angle at the exit end.
Suppose that the matrix $\mathbf{X}$ connects the electronic wave function
at the input end with Eq. (\ref{WAVE00}) and that at the exit end with (\ref%
{WAVEdd}), so that we can connect the input and output wave functions by the
following equation%
\begin{equation}
\left( 
\begin{array}{c}
\psi _{A}(x_{e}) \\ 
\psi _{B}(x_{e})%
\end{array}%
\right) =\mathbf{X}\left( 
\begin{array}{c}
\psi _{A}(0) \\ 
\psi _{B}(0)%
\end{array}%
\right) ,  \label{WAVE00dd}
\end{equation}%
with%
\begin{equation}
\mathbf{X}\mathbf{=}\left( 
\begin{array}{cc}
x_{11} & x_{12} \\ 
x_{21} & x_{22}%
\end{array}%
\right) =\dprod\limits_{j=1}^{N}M_{j}(w_{j},E,k_{y}).  \label{XXmatrix}
\end{equation}%
By substituting Eqs. (\ref{WAVE00},\ref{WAVEdd}) into Eq. (\ref{WAVE00dd}),
we have the following relations%
\begin{eqnarray}
t &=&(1+r)x_{11}+(e^{i\theta _{0}}-re^{-i\theta _{0}})x_{12},  \label{rt1} \\
te^{i\theta _{e}} &=&(1+r)x_{21}+(e^{i\theta _{0}}-re^{-i\theta _{0}})x_{22}.
\label{rt2}
\end{eqnarray}%
Sovling the above two equations, we find the reflection and transmission
coefficients given by%
\begin{eqnarray}
r(E,k_{y}) &=&\frac{(x_{22}e^{i\theta _{0}}-x_{11}e^{i\theta
_{e}})-x_{12}e^{i(\theta _{e}+\theta _{0})}+x_{21}}{(x_{22}e^{-i\theta
_{0}}+x_{11}e^{i\theta _{e}})-x_{12}e^{i(\theta _{e}-\theta _{0})}-x_{21}},
\label{rrcoeff} \\
t(E,k_{y}) &=&\frac{2\cos \theta _{0}}{(x_{22}e^{-i\theta
_{0}}+x_{11}e^{i\theta _{e}})-x_{12}e^{i(\theta _{e}-\theta _{0})}-x_{21}},
\label{ttcoeff}
\end{eqnarray}%
where we have used the property of $\det [\mathbf{X}]=1$. With Eqs. (\ref%
{PropagationAB}), (\ref{PisiA0}) and (\ref{PisiB0}), now we are able to
calculate the two components of the electronic wave function at any position
as follows%
\begin{eqnarray}
\psi _{A}(x) &=&\psi _{i}(E,k_{y})\left[ \left( 1+r\right)
Q_{11}+(e^{i\theta _{0}}-re^{-i\theta _{0}})Q_{12}\right] ,  \label{PisiAAx}
\\
\psi _{B}(x) &=&\psi _{i}(E,k_{y})\left[ \left( 1+r\right)
Q_{21}+(e^{i\theta _{0}}-re^{-i\theta _{0}})Q_{22}\right] ,  \label{PisiBBx}
\end{eqnarray}%
where $Q_{mn}$ are elements of matrix $Q$. When we consider the translation
of the electron in the $y$ direction for obtaining the wave functions $%
\tilde{\psi}_{A,B}(x,y),$ the above functions $\psi _{A,B}(x)$ have to be
producted by the factor, $Y(y,k_{y})=\exp \left( ik_{y}\Delta y_{j}\right)
\tprod\limits_{i=1}^{j-1}\exp (ik_{y}\Delta y_{i})$, where $\Delta
y_{i}=w_{i}\tan \theta _{i}$ and $\Delta y_{j}=\Delta x_{j}\tan \theta _{j}$%
. Therefore we finally get%
\begin{eqnarray}
\tilde{\psi}_{A}(x,y) &=&\psi _{i}(E,k_{y})Y(y,k_{y})\left[ \left(
1+r\right) Q_{11}+(e^{i\theta _{0}}-re^{-i\theta _{0}})Q_{12}\right] , \\
\tilde{\psi}_{B}(x,y) &=&\psi _{i}(E,k_{y})Y(y,k_{y})\left[ \left(
1+r\right) Q_{21}+(e^{i\theta _{0}}-re^{-i\theta _{0}})Q_{22}\right] .
\end{eqnarray}%
We emphasize that these equations are very useful to fully describe the
evolutions of the two-component pseudospinor's wave function inside the
potential or barrier structures of graphene when the incident electron's
wavepacket $\psi _{i}(E,k_{y})$ is given; and furthermore all these formulae
are also suitable for multi-potential structures in graphene, not only for
periodic potential structures. In the following discussions, we will discuss
the properties of the electronic band structure and transmission for the
graphene-based periodic squared potential structures.

\section{Results and Discussions}

In this section, we would like to discuss some unique properties of the band
structures in the graphene-based periodic-potential systems by using the
above transfer method. First, let us invetigate the electron's bandgap for
an infinite periodic structure $(AB)^{N}$, where the periodic number $N$
tends to infinity. The magnitude and width of the potential $A$ ($B)$ are
with the electrostatic potential $V_{A(B)}$ and width $w_{A(B)}$, as shown
in Fig. 1. According to the Bloch's theorem, the electronic dispersion at
any incident angle follows the relation%
\begin{eqnarray}
2\cos [\beta _{x}D] &=&Tr[M_{A}M_{B}]  \notag \\
&=&2\cos [q_{A}w_{A}+q_{B}w_{B}]+\frac{[2\cos (\theta _{A}-\theta _{B})-2]}{%
\cos \theta _{A}\cos \theta _{B}}\sin (q_{A}w_{A})\sin (q_{B}w_{B}).
\label{BG100}
\end{eqnarray}%
When the incident energy of the electron satisfies $V_{B}<E<V_{A}$, we have $%
\theta _{A}<0$, $q_{A}<0$, $\theta _{B}>0$, and $q_{B}>0$ for the
propagating modes. Then if $q_{A}w_{A}=-q_{B}w_{B}$, the above equation (\ref%
{BG100}) becomes%
\begin{equation}
\cos [\beta _{x}D]=1+\frac{[1-\cos (2\theta _{A})]}{\cos ^{2}\theta _{A}}%
|\sin (q_{A}w_{A})|^{2}.  \label{BG111}
\end{equation}%
This equation indicates that, when $q_{A}w_{A}=-q_{B}w_{B}\neq m\pi $ and $%
\theta _{A}\neq 0$, there is no real solution for $\beta _{x}$, i.e.,
exsiting a bandgap; Note this bandgap will be close at normal incident case (%
$\theta _{A}=0$) from Eq. (\ref{BG111}). Therefore, the location of the
touching point of the bands is exactly given by $q_{A}w_{A}=-q_{B}w_{B}$ at $%
\theta _{A}=0$, i. e.,$\ k_{A}w_{A}=-k_{B}w_{B}$ or $%
(V_{A}-E)w_{A}=(E-V_{B})w_{B}$.

Figure 2 shows clearly that a band gap opens exactly at energy $E=25$meV
under the inclined incident angles (i.e., $k_{y}\neq 0$), where the
condition $q_{A}d_{A}=-q_{B}d_{B}\neq m\pi $ is satisfied. At the case of
normal incidence ($\theta _{A}=\theta _{B}=0$), the upper and lower bands
linearly touch together and form a new double-cone Dirac point. The location
of the new Dirac point is governed by the equality: $%
(V_{A}-E)w_{A}=(E-V_{B})w_{B}$. For the graphene-based periodic-barrier
structure with $V_{A}\neq 0$ and $V_{B}=0$, the new Dirac point is exactly
located at $E=V_{A}/(1+w_{B}/w_{A})$. From Fig. 2(a-c), one can also find
that the location of the new Dirac point is independent of the lattice
constants; and the position of the the opened gap associated with the new
Dirac point is not only independent of the lattice constants but also is
weakly dependent on the incident angles [also see the transmission curves in
Figs. 3a and 3b for the finite structures, for example, $(AB)^{25}$]; while
other bandgap structuers are not only dependent on the lattice contants but
also strongly dependent on different angles (i. e., different $k_{y}$). The
properties of the opened gap associated with the new Dirac point are very
similar to that in the one-dimensional photonic crystals containing the
left-haned materials \cite{Li2003}, where the so-called zero (volumn)
averaged index gap is independent of the lattice constant but only dependent
on the ratio of the thicknesses of the right- and left-handed materials \cite%
{Li2003}. From Fig. 4, one can also find that the locations of both the new
Dirac point and the opened gap are determined by the ratio value of $%
w_{A}/w_{B}$ for the cases with the fixed heights $V_{A}$ of the potentials.
From the above discussion, we find that the volume-averaged wavenumber at
the energy of the new Dirac point is zero, therefore such a opened gap
associated with the new Dirac point may be called as the zero-averaged
wavenumber gap.

From Figs. 2(a-c) and 5(a-c), one can also noted that the slope of the band
edges near the new Dirac point gradually becomes smaller as the lattice
constant increases under the fixed ratio $w_{A}/w_{B}$ and the fixed
potential height, and such phenomena have been pointed out in recent work by
Brey and Fertig \cite{Brey2009}. Actually, from Eq. (\ref{BG100}), we can
see that as the values of $q_{A}w_{A}$ and $-q_{B}w_{B}$ gradually reach the
values of $m\pi $ ($m=1,2,3,\cdots $), the slope of the band edges near the
new Dirac point becomes smaller [see Figs. 2(a-c) and 5(a-c)]; once the
condition $q_{A}w_{A}=-q_{B}w_{B}=m\pi $ is satisfied, the zero-averaged
wavenumber gap will be close and a pair of new zero-averaged wavenumber
states emerges from $k_{y}=0$ [see Fig. 5(c)]. Here we would like to
emphasize that the properties of these novel zero-averaged wavenumber states
are similar to that of the zero-energy states in the previous work \cite%
{Brey2009}. Figure 5(d) shows how the zero-averaged wavenumber gap does
gradually close with the increasing of the lattice constant under the fixed
transversal wavenumber $k_{y}=0.01$nm$^{-1}$ and the fixed ratio $%
w_{A}/w_{B} $. From Fig. 5(d), one can find that the zero-averaged
wavenumber gap is very speical and it is open and close oscillationly with
the increasing of the lattice constant, and the center position of the
zero-averaged wavenumber gap is also independent of the lattice constant.
However, the other opened gaps are largely shifted with the increasing of
the lattice constant.

Now we turn to consider the transmission of an electron passing through a
finite graphene-based periodic-potential structure, e. g., $(AB)^{30}$, with
the width deviation under different incident angles. Figure 6 shows the
effect of the structural disorder on electronic transimitivities. Figures
6a, 6b and 6c correspond to the transmissions through the periodic-potential
structures with the width deviation (random uniform deviate) of $\pm 2.5$nm, 
$\pm 3.75$nm, and $\pm 5.0$nm, respectively. From Figs. (6a)-(6c), it is
clear that the higher opened gap is destroyed by strong disorder, but the
zero-averaged wavenumber gap survives. The robustness of the zero-averaged
wavenumber gap comes from the fact that the zero-averaged wavenumber
solution remains invariant under disorder that is symmetric ($+$ and $-$
equally probable). It should be emphasized again that the position of the
zero-averaged wavenumber gap near the new Dirac point is insensitive to both
the incident angles and the disorder. [see Fig. 3b and Fig. 6].

Finally, we turn our attention to the effect of a defect potential on the
property of the electron's transport inside the zero-averaged wavenumber
gap. Here we consider the transmission of an electron passing through a
graphene-based periodic potential structure with a defect potential, e. g., $%
(AB)^{30}D(BA)^{30},$ where the symbol "D" denotes the defect barrier. In
Fig. 7(a), it shows the electronic transmitivity through the structure under
different incident angles. One can see that there is a defect mode
respectively occuring inside the zero-averaged wavenumber gap and the higher
bandgap, and the location of the defect mode inside the zero-averaged
wavenumber gap is very weakly dependent on the incident angle but the defect
mode in the higher bandgap is strongly sensitive to the incident angle. It
is clear shown in Fig. 7(b) that the energy of the defect mode inside the
zero-averaged wavenumber gap is almost independent of the angles while the
location of the defect mode in the higher bandgap has a large shift with the
increasing of the incident angle.

\section{Conclusions}

In summary, we studied the electronic band structures and transmission of
carriers in the graphene-based 1D periodic squared-potential superlattices.
For the 1D periodic squared-potential structure, it is found that a new
Dirac point does exactly occurs at the energy that corresponds to the zero
(volume) averaged wavenumber inside the system, and the location of such a
new Dirac point is independent of lattice constants but dependent on the
ratio of potential widths. It is also shown that the location of the
zero-averaged wavenumber gap associated with the new Dirac point is not only
independent of lattice constants but is also weakly dependent on the
incident angles. As the lattice constant increases for the structures with
the fixed ratio of potential widths, this zero-averaged wavenumber gap is
open and close oscillationaly around the energy of the new Dirac point. We
further showed the robustness of the zero-averaged wavenumber gap against
the structural disorder, which has a sensitive effect on other opened gaps
of the system. Finally we saw that the defect mode inside the zero-averaged
wavenumber gap is weakly dependent on the insident angles while the defect
mode in other gaps are highly dependent on the incident angles. Our
analytical and numerical results on the properties of the new Dirac point,
the novel band gap structure and defect mode are hopefully of use to
pertinent experiments.

\begin{acknowledgments}
This work is supported by RGC 403609 and CUHK 2060360, and partially
supported by the National Natural Science Foundation of China (10604047) and
HKUST3/06C.
\end{acknowledgments}

\newpage

\begin{center}
{\Huge Figures Captions:}
\end{center}

Fig. 1. (a) Schematic representation of the finite periodic squared
potential structure in $x-y$ plane. Dark regions denote the electrodes to
apply the periodic potentials on the graphene. (b) The profiles of the
periodic potentials applied on the monolayer graphene.

Fig. 2. (Color online). Electronic band structures for (a) $w_{A}=w_{B}=20$%
nm, (b) $w_{A}=w_{B}=30$nm and (c) $w_{A}=w_{B}=40$nm, with $V_{A}=50$meV
and $V_{B}=0$ in all cases. The dashed lines denote the "light cones" of the
incident electrons, and the dot line denotes the location of the new Dirac
points.

Fig. 3. (Color online). Transmitivities of the finite periodic-potential
structure $(AB)^{25}$ under (a) different lattice contants with a fixed
ratio $w_{A}/w_{B}=1$ and an incident angle $\theta _{0}=10^{\circ }$ and
(b) different incident angles with the fixed lattice parameters $%
w_{A}=w_{B}=30$nm.

Fig. 4. (Color online). Electronic band structures for (a) $w_{A}/w_{B}=1$,
(b) $w_{A}/w_{B}=3/2$, and (c) $w_{A}/w_{B}=2$, with $V_{A}=50$meV, $V_{B}=0$
and $w_{B}=20$nm in all cases. The dashed lines denote the locations of the
new Dirac points.

Fig. 5. (Color online). Electronic band structures for (a) $w_{A}=w_{B}=60$%
nm, (b) $w_{A}=w_{B}=80$nm and (c) $w_{A}=w_{B}=100$nm; and (d) dependence
of the bandgap structure on the lattice constant $w$ with a fixed
transversal wavenumber $k_{y}=0.01$nm$^{-1}$ and $w_{A}=w_{B}=w$. Other
parameters are the same as in Fig. 2.

Fig. 6. (Color online). The effect of the structural disorder on electronic
transimitivity $T=|t|^{2}$ under different incident angles: the solid lines
for the incident angle $\theta _{0}=1^{\circ }$, the dashed lines for $%
\theta _{0}=5^{\circ },$the dashed-dot lines for $\theta _{0}=10^{\circ }$,
and the short-dashed lines for $\theta _{0}=15^{\circ }$, with $V_{A}=50$meV
and $V_{B}=0,$ and $w_{A}=w_{B}=(20+R)$nm, where $R$ is a random number for
the case (a) between +2.5 nm and -2.5 nm, for the case (b) between +3.75 nm
and -3.75 nm, and for the case (c) between +5 nm and -5 nm.

Fig. 7. (Color online). (a) The transmitivity as a function of the incident
electronic energy in a periodic-potential structure with a defect potential $%
D$, $(AB)^{30}D(BA)^{30}$. (b) Dependence of the defect modes on the
incident angles. The parameters of the defect potential are $w_{D}=70$nm and 
$V_{D}=50$meV, and other parameters are $V_{A}=50$meV$,$ $V_{B}=0,$and $%
w_{A}=$ $w_{B}=20$nm$.$

\newpage

%%%%%%%%%%%%%%%%%%FIG1%%%%%%%%%%%%%
\begin{figure}[b]
\centering
\includegraphics[width=8cm]{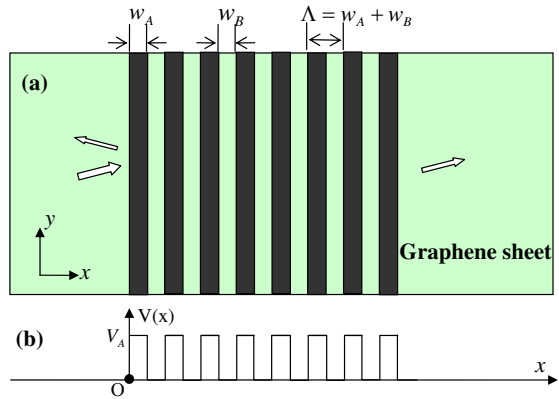}
\caption{(Color online) (a) Schematic representation of the finite periodic
squared potential structure in $x-y$ plane. Dark regions denote the
electrodes to apply the periodic potentials on the graphene. (b) The
profiles of the periodic potentials applied on the monolayer graphene.}
\label{fig:FIG1}
\end{figure}
%%%%%%%%%%%%%%%%%%%%%%%%%%%%%%%%%%%%

\newpage

%%%%%%%%%%%%%%%%%%FIG2%%%%%%%%%%%%%
\begin{figure}[b]
\centering
\includegraphics[width=10cm]{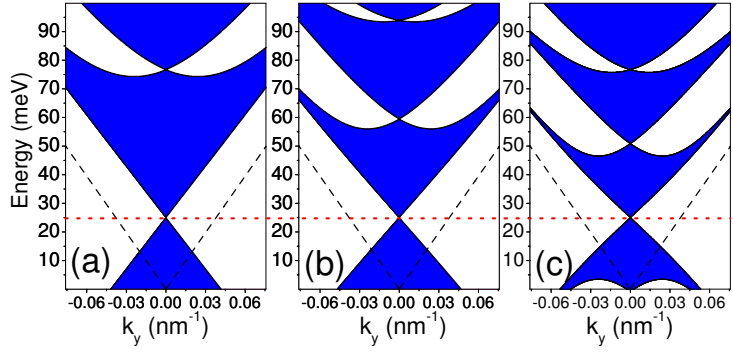}
\caption{(Color online) Electronic band structures for (a) $w_{A}=w_{B}=20$%
nm, (b) $w_{A}=w_{B}=30$nm and (c) $w_{A}=w_{B}=40$nm, with $V_{A}=50$meV
and $V_{B}=0$ in all cases. The dashed lines denote the "light cones" of the
incident electrons, and the dot line denotes the location of the new Dirac
points. }
\label{fig:FIG2}
\end{figure}
%%%%%%%%%%%%%%%%%%%%%%%%%%%%%%%%%%%%

%%%%%%%%%%%%%%%%%%FIG3%%%%%%%%%%%%%
\begin{figure}[b]
\centering
\includegraphics[width=8cm]{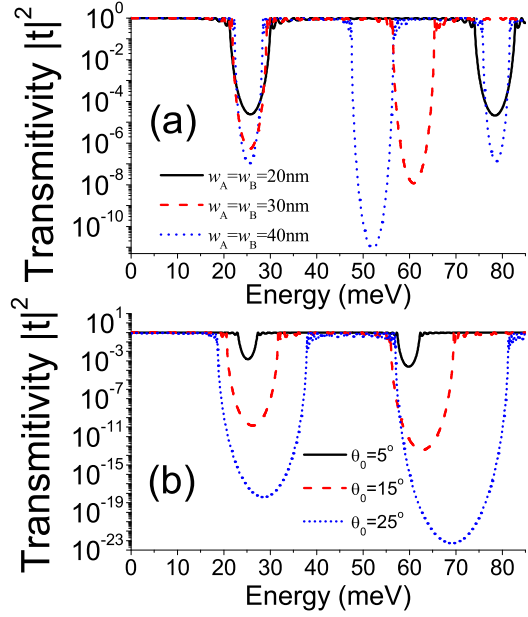}
\caption{(Color online) Transmitivities of the finite periodic-potential
structure $(AB)^{25}$ under (a) different lattice contants with a fixed
ratio $w_{A}/w_{B}=1$ and an incident angle $\protect\theta _{0}=10^{\circ }$
and (b) different incident angles with the fixed lattice parameters $%
w_{A}=w_{B}=30$nm. }
\label{fig:FIG3}
\end{figure}
%%%%%%%%%%%%%%%%%%%%%%%%%%%%%%%%%%%%

%%%%%%%%%%%%%%%%%%FIG4%%%%%%%%%%%%%
\begin{figure}[b]
\centering
\includegraphics[width=8cm]{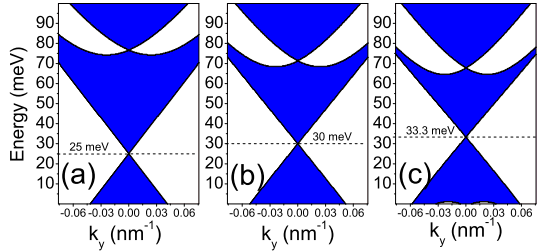}
\caption{(Color online) Electronic band structures for (a) $w_{A}/w_{B}=1$,
(b) $w_{A}/w_{B}=3/2$, and (c) $w_{A}/w_{B}=2$, with $V_{A}=50$meV, $V_{B}=0$
and $w_{B}=20$nm in all cases. The dashed lines denote the locations of the
new Dirac points. }
\label{fig:FIG4}
\end{figure}
%%%%%%%%%%%%%%%%%%%%%%%%%%%%%%%%%%%%

%%%%%%%%%%%%%%%%%%FIG5%%%%%%%%%%%%%
\begin{figure}[b]
\centering
\includegraphics[width=8cm]{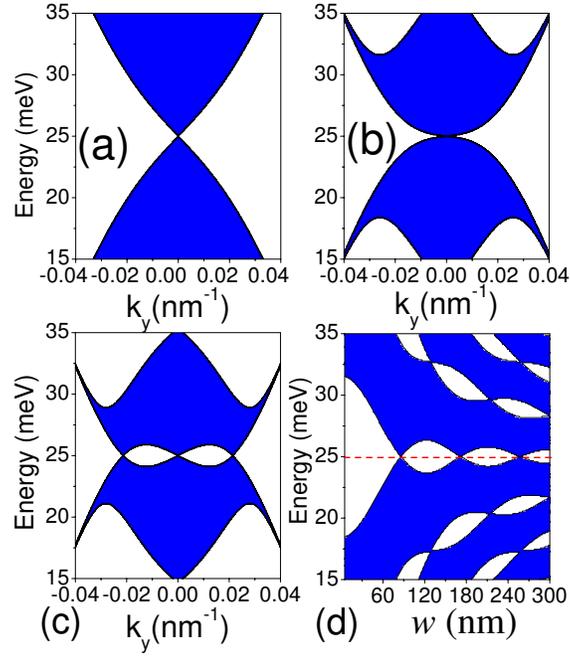}
\caption{(Color online) Electronic band structures for (a) $w_{A}=w_{B}=60$%
nm, (b) $w_{A}=w_{B}=80$nm and (c) $w_{A}=w_{B}=100$nm; and (d) dependence
of the bandgap structure on the lattice constant $w$ with a fixed
transversal wavenumber $k_{y}=0.01$nm$^{-1}$ and $w_{A}=w_{B}=w$. Other
parameters are the same as in Fig. 2. }
\label{fig:FIG5}
\end{figure}
%%%%%%%%%%%%%%%%%%%%%%%%%%%%%%%%%%%%

%%%%%%%%%%%%%%%%%%FIG6%%%%%%%%%%%%%
\begin{figure}[b]
\centering
\includegraphics[width=8cm]{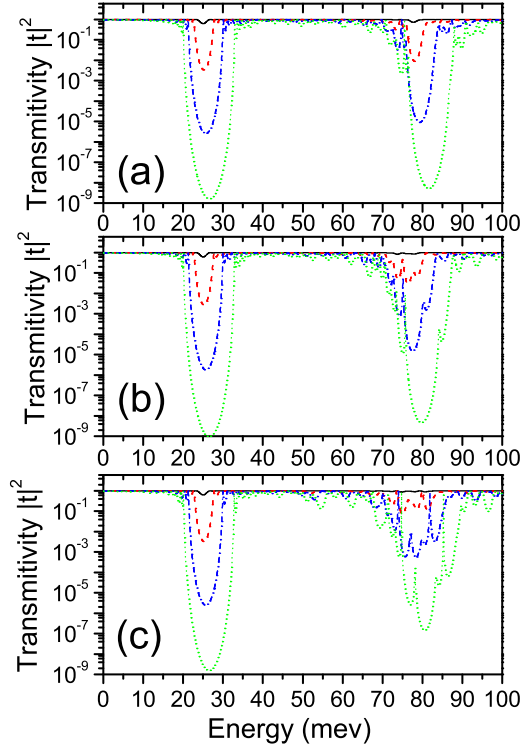}
\caption{(Color online) The effect of the structural disorder on electronic
transimitivity $T=|t|^{2}$ under different incident angles: the solid lines
for the incident angle $\protect\theta _{0}=1^{\circ }$, the dashed lines
for $\protect\theta _{0}=5^{\circ },$the dashed-dot lines for $\protect%
\theta _{0}=10^{\circ }$, and the short-dashed lines for $\protect\theta %
_{0}=15^{\circ }$, with $V_{A}=50$meV and $V_{B}=0,$ and $w_{A}=w_{B}=(20+R)$%
nm, where $R$ is a random number for the case (a) between +2.5 nm and -2.5
nm, for the case (b) between +3.75 nm and -3.75 nm, and for the case (c)
between +5 nm and -5 nm. }
\label{fig:FIG6}
\end{figure}
%%%%%%%%%%%%%%%%%%%%%%%%%%%%%%%%%%%%

%%%%%%%%%%%%%%%%%%FIG7%%%%%%%%%%%%%
\begin{figure}[b]
\centering
\includegraphics[width=8cm]{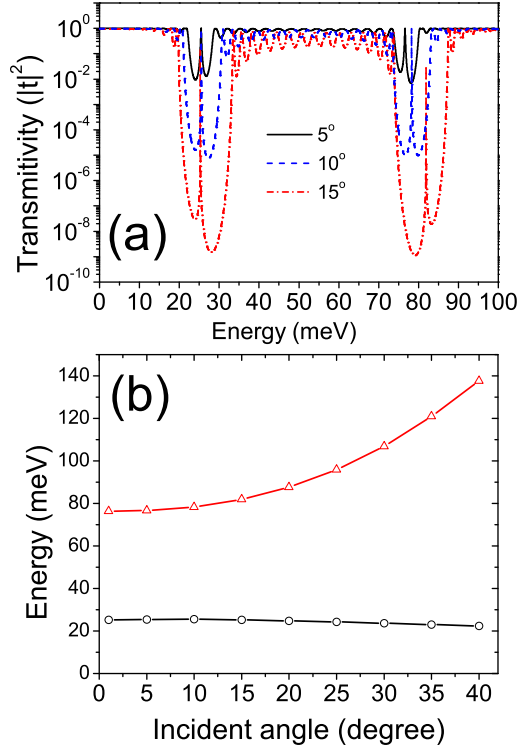}
\caption{(Color online) (a) The transmitivity as a function of the incident
electronic energy in a periodic-potential structure with a defect potential $%
D$, $(AB)^{30}D(BA)^{30}$. (b) Dependence of the defect modes on the
incident angles. The parameters of the defect potential are $w_{D}=70nm$ and 
$V_{D}=50meV$, and other parameters are $V_{A}=50meV,$ $V_{B}=0,$and $w_{A}=$
$w_{B}=20nm.$ }
\label{fig:FIG7}
\end{figure}
%%%%%%%%%%%%%%%%%%%%%%%%%%%%%%%%%%%%

\end{document}